\documentclass{sig-alt-release2}
\usepackage{amsmath}
\usepackage{dsfont}
\usepackage{algorithm}
\usepackage{algorithmic}
\usepackage{array}
\usepackage{subfigure}
\usepackage{latexsym}
\usepackage{amsmath}
\usepackage{amssymb}
\usepackage{bm}
\usepackage{graphicx}
\usepackage{wrapfig}
\usepackage{fancybox}

\newcommand{\remove}[1]{}
\newcommand{\palpha}{p_\alpha}
\newcommand{\qiarrow}{q_{i\curvearrowright}}
\newcommand{\qarrow}{q_{\curvearrowright}}
\newcommand{\Q}{\mathcal{Q}}
\newcommand{\Ptoi}{\mathcal{P}^i}

\newcommand{\Xtoi}{\mathcal{X}^i}
\newcommand{\PplusQ}{\qiarrow+\sum_{\beta\in i}p_\beta}
\newcommand{\PplusQalpha}{\qiarrow+\sum_{\alpha\in i}p_\alpha}
\newcommand{\piarrow}{p^i_{\circlearrowright}}
\newcommand{\djalpha}{d_{j}^{~\alpha}}
\newcommand{\xvalpha}{\textbf{x}^{\alpha}}
\newcommand{\xalpha}{x^{\alpha}}
\newcommand{\xmodule}{x^{(i)}}
\newcommand{\pmodule}{p^{(i)}}

\usepackage{color}

\begin{document}
\title{Partitioning Networks with Node Attributes by Compressing Information Flow}
\numberofauthors{4} %  in this sample file, there are a *total*
\author{
\alignauthor
Laura M. Smith\\
       \affaddr{ Department of Mathematics}\\
       \affaddr{California State University}\\
       \affaddr{Fullerton, CA }\\
       \email{lausmith@fullerton.edu}
\alignauthor Linhong Zhu \\
       \affaddr{Information Sciences Institute}\\
       \affaddr{U. of Southern California}\\
       \affaddr{Marina del Rey, CA 90292}\\
       \email{linhong@isi.edu}
\alignauthor
Kristina Lerman\\
       \affaddr{Information Sciences Institute}\\
       \affaddr{U. of Southern California}\\
       \affaddr{Marina del Rey, CA 90292}\\
       \email{lerman@isi.edu}
  % use '\and' if you need 'another row' of author names
  \and
\alignauthor
Allon G. Percus\\
       \affaddr{Claremont Graduate U.}\\
       \affaddr{Claremont, CA 91711}\\
      \email{allon.percus@cgu.edu}
}
\date{}

\maketitle
\begin{abstract}

Real-world networks are often organized as modules or communities of similar nodes that serve as functional units. These networks are also rich in content, with nodes having distinguishing features or attributes. In order to discover a network's modular structure, it is necessary to take into account not only its links but also node attributes.
We describe an information-theoretic method that identifies modules by compressing descriptions of information flow on a network. Our formulation introduces node content into the description of information flow, which we then minimize to discover groups of nodes with similar attributes that also tend to trap the flow of information.
The method has several advantages: it is conceptually simple and does not require ad-hoc parameters to specify the number of modules or to control the relative contribution of links and node attributes to network structure.
We apply the proposed method to partition real-world networks with known community structure. We demonstrate that adding node attributes helps recover the underlying community structure in content-rich networks more effectively than using links alone. In addition, we show that our method is faster and more accurate than alternative state-of-the-art algorithms.

\end{abstract}


\section{Introduction}
One of the fundamental tasks in network analysis is to partition a network into clusters, or modules, of similar nodes, which often correspond to functional units in biological networks~\cite{Ravasz2002Hierarchical,Rives03} or communities in social networks~\cite{Newman2006}. The vast majority of methods developed for this task rely on network topology, i.e., the structure of links between nodes, and treat the nodes themselves as indistinguishable. For example, spectral partitioning methods~\cite{Chung:Spectral:97,spectral-tutorial,Spielman07} identify which links to cut to separate the network into disconnected components, while modularity-based approaches~\cite{Newman2006,Fortunato10} find clusters of densely connected nodes. Real-world networks, however, are often rich in content, with nodes that have distinguishing features or attributes. Individuals in a social network differ in age, gender, education and interests, while articles in a scientific paper citation network have different words and topics. The similarities and differences in the content of nodes can affect the patterns of linking, particularly in social networks~\cite{homophily,Kang12homophily}, and taking them into account may improve the quality of the discovered modules. This observation has inspired several attempts to partition content-rich networks~\cite{icde/QiAH12,BACG12,wwwRuanFP13,YangKDD09,zhouCY09,ZhuNC11,getoor13}. In contrast to these works, we describe a parameter-free, conceptually simple method that combines information in links and node attributes to partition a network.

% Allon's
\remove{Our method is situated in the context of Rosvall $\&$ Bergstrom's novel information-theoretic framework \cite{Rosvall08}}
% Our method builds upon the work of Rosvall $\&$ Bergstrom \cite{Rosvall08}, who proposed a novel information-theoretic framework
Our method is situated in the information-theoretic framework introduced by Rosvall $\&$ Bergstrom~\cite{Rosvall08}
for finding the modular structure of networks.  Their approach is inspired by an observation that information flows on a network tend to get trapped within modules. As a consequence, it is possible to compress the description of information flow by reusing names of nodes in different modules. Using random walks as a proxy for information flow, their method partitions the network so as to minimize the Map Equation, which gives the expected description length of a random walk. Thus, the approach exploits the duality between identifying structure and the compression problem to identify the optimal number of modules in the network and to assign the nodes to modules.

{To describe the flow of information in a content-rich network,
however, it is not sufficient to account for the node names and modules.
We need an effective means of accounting for node attributes as well.}
% For describing the flow of information on a content-rich network, in addition to names of nodes and modules, we also need to account for node attributes.
To this end, we introduce the \emph{Content Map Equation}, which incorporates node attributes into a description of information flow, and use it to compress the flow of information on content-rich networks. The Content Map Equation groups nodes into modules not only when information frequently flows between them, but also when they have similar attributes.

Our method has several desirable properties. First, it is conceptually simple and treats links and attributes on an equal footing.
%it prefers to group nodes with the same attributes in the same modules.
It is parameter-free and does not require us to specify the number of modules ahead of time. It is not sensitive to content representation, i.e., how many attributes are used to characterize nodes.
{Additionally, it does not require a parameter to control the
relative contributions of links and attributes in encoding network
information.  This is contrast to other
methods~\cite{icde/QiAH12,wwwRuanFP13,BACG12,zhouCY09,getoor13}, whose
quality relies on successfully tuning such a parameter.}
% Additionally, since both links and attributes contribute equally to encoding the information in a network, it does not require a parameter to control the contribution of each, in contrast to other methods~\cite{icde/QiAH12,wwwRuanFP13,BACG12,zhouCY09,getoor13}.

Finding a minimum solution to the Content Map Equation is in most cases a hard optimization problem.
% Following
{Similarly to} Rosvall \& Berg\-strom, we use  a greedy bottom-up search to find a locally optimal solution. In that procedure, each node starts in its own module, and the search proceeds by merging modules so as to minimize the total description length. However, this becomes intractable for large networks. To address this problem, we propose a top-down search strategy that has better scaling properties than the original greedy algorithm. We show that it leads to dramatically better computational performance without sacrificing result quality.

We use the proposed method to partition several real-world networks with node attributes and a known community structure. We demonstrate that the Content Map Equation identifies better modules than the original Map Equation, which does not use content information. We also show that our method outperforms alternative methods that use both links and attributes, both in terms of runtime and in terms of the quality of the discovered modules.

% contributions
In the rest of the paper, we first review related work (Section~\ref{sec:background}), including Rosvall \& Berg\-strom's Map Equation. In Section~\ref{sec:content} we introduce the Content Map Equation that includes node attributes in a description of information in a network. We illustrate on toy networks the difference in the resulting partitions. In Section~\ref{sec:greedy}, we describe a greedy bottom-up algorithm that uses the Content Map Equation to minimize the description length of a random walk.  The  bottom-up algorithm does not scale to large networks; therefore, we propose a top-down algorithm with random restarts that significantly speeds up the compression problem. In Section~\ref{sec:results} we use the proposed methods to partition real-world networks with known community structure and demonstrate that our algorithm is faster and more accurate than competing methods.

\section{Background and Related Work}\label{sec:background}
\remove{
\begin{table*}[!t]
  \centering
  \caption{Summary of related work on community detection using both links and node attributes. $n$: number of nodes, $l$: number of edges, $d$: number of attributes, $k$: number of iterations, $m$: number of communities, and $\delta_n$/$\delta_l$/$\delta_d$: a small number of nodes/edges/attributes in the neighborhood of a node.}\label{tab:relatedcompare}
  \begin{tabular}{|c|c|c|c|c|c|c|c|c|c|c|c|c|c|c|}
    \hline
Method&\cite{SDM12}&\cite{Entropypartition}&\cite{HendersonEPF10}&\cite{LongICML2006}
&\cite{icde/QiAH12}&\cite{wwwRuanFP13}&\cite{BACG12}&\cite{CESNAICDM13}
&\cite{YangKDD09}&\cite{zhouCY09}&\cite{ZhuNC11}&\cite{getoor13}&B-CME&T-CME\\
\hline
Generative Model&&&$\surd$&
&&&$\surd$&$\surd$
&$\surd$&&&$\surd$&&\\
\hline
Matrix factorization&&&&$\surd$
&$\surd$&&&
&&&&&&\\
\hline
Information Theory&$\surd$&$\surd$&
&&&&
&&&&&&$\surd$&$\surd$\\
\hline
Hybrid&&&&
&&$\surd$&&
&&$\surd$&$\surd$&&&\\
\hline
Multi-attributes&$\surd$&$\surd$&$\surd$&$\surd$&$\surd$&$\surd$&&$\surd$&&&$\surd$&$\surd$&$\surd$&$\surd$\\
\hline
Parameter-free&$\surd$&$\surd$&&
&$\surd$&&&&&&$\surd$&&$\surd$&$\surd$\\
\hline
$O(km^2(n+l+d))$ and above&$\surd$&?&$\surd$&$\surd$
&?&?&?&
&$\surd$&$\surd$&&&$\surd$&\\
\hline
$O(kmn(\delta_n+\delta_l+\delta_d))$&&&&
&&&&$\surd$
&&&$\surd$&$\surd$&&$\surd$\\
\hline
  \end{tabular}
\end{table*}
}
\begin{table*}[!t]
  \centering
  \caption{Summary of related work on community detection using both links and node attributes. $n$: number of nodes, $l$: number of links, $d$: number of attributes, $k$: number of iterations, $m$: number of communities, and $\delta_n$/$\delta_l$/$\delta_d$: number of nodes/links/attributes in the neighborhood of a node. $MF$ and $IR$ stand for matrix factorization and information theory-based approaches. The last two columns list features of the methods proposed in this paper. } \label{tab:relatedcompare}
  \begin{tabular}{|@{\ }c@{\ }|c|c|c|c|c|c|c|c|c|c|c|c||c|c|}
    \hline
\emph{Method} & \multicolumn{5}{c|}{\emph{Generative}} &	\multicolumn{3}{c|}{\emph{Hybrid}} &	 \multicolumn{2}{c|}{\emph{MF}} &	\multicolumn{2}{c|}{\emph{IR}} &	 \multicolumn{2}{|c|}{\emph{Proposed}}			\\
\cline{2-15}
&	\cite{HendersonEPF10}	&	\cite{BACG12}	&	\cite{CESNAICDM13}	&	\cite{YangKDD09}	&	 \cite{getoor13}	&	\cite{wwwRuanFP13}	 &	\cite{zhouCY09}	&	\cite{ZhuNC11}	&	\cite{LongICML2006}	 &	\cite{icde/QiAH12}	&	\cite{SDM12}	&	\cite{Entropypartition}	&	 \textsc{b-cme}	&	\textsc{t-cme}	\\ \hline
Multi-attribute	&	$\surd$	&		&	$\surd$	&		&	$\surd$	&	$\surd$	&		&	$\surd$	&	$\surd$	&	 $\surd$	&	$\surd$	&	$\surd$	&	 $\surd$	&	$\surd$	\\
Parameter-free	&		&		&		&		&		&		&		&	$\surd$	&		&	$\surd$	&	$\surd$	&	$\surd$	 &	$\surd$	&	$\surd$	\\
$O(km^2(n+l+d))$ and above	&	$\surd$	&	?	&		&	$\surd$	&		&	?	&	$\surd$	&		&	$\surd$	&	 ?	&	$\surd$	&	?	&	$\surd$	&		 \\
$O(kmn(\delta_n+\delta_l+\delta_d))$	&		&		&	$\surd$	&		&	$\surd$	&		&		&	$\surd$	&		 &		&		&		&		&	$\surd$	 \\
\hline
  \end{tabular}
\end{table*}

Recently, there has been an explosion of interest in community detection using both links and node attributes. Proposed techniques range from generative modeling, to matrix factorization, to information theoretic approaches. Our (non-exhaustive) summary is shown in Table~\ref{tab:relatedcompare}.

Most of the existing generative modeling approaches, such as~\cite{YangKDD09,HendersonEPF10,BACG12,getoor13}, extend the mixed-membership model~\cite{FuICML09} with the assumption that communities and attributes together generate links. In contrast, Yang et al.~\cite{CESNAICDM13} assume that communities ``generate" both links and attributes, and propose an alternative way to combine content information using probabilistic modeling. However, in practice this approach only supports nodes with single-dimensional attributes due to the embedded logistic modeling, while all remaining approaches using generative modeling in Table~\ref{tab:relatedcompare} support multi-dimensional attributes.

Another popular category for community detection methods using both links and content is the hybrid approach~\cite{wwwRuanFP13,zhouCY09,ZhuNC11}. The general workflow of the hybrid approach is as follows: it first generates content links based on attribute vector similarity, and then combines content links with topological links to detect communities.

Compared to generative modeling and hybrid approaches, fewer methods have been developed that use matrix factorization or information theory. Matrix factorization~\cite{LongICML2006,icde/QiAH12} aims to jointly co-factorize the adjacency matrix of the graph and the node-attribute matrix to obtain the low-ranked node-community matrix.

From the information theoretic view, the entropy-based approach~\cite{Entropypartition} aims to detect communities with low entropy and high modularity. Akoglu~\cite{SDM12} extracts cohesive subgraphs by compressing the storage cost of matrices.

In our work, we approach the problem of partitioning content-rich networks from another information theoretic perspective: % compressing the flow of information on the network. %This is different from both generative modeling and compressing the storage cost of matrices. Generative modeling assumes that communities and attributes together generate the links, while this work assumes that information flow forms new links~\cite{Narang13snakdd}.
exploiting the duality between identifying communities and compressing information, which differs from the matrix storage compression~\cite{SDM12}.
Our method is inspired by Rosvall \& Bergstrom~\cite{Rosvall08}, who proposed compressing information flows on a network in order to identify modules. Using random walks as a proxy for information flow, their method compresses the description length of a random walk by minimizing the \emph{Map Equation}.  Through this optimization, communities emerge as modules with large internal information flows form.
{We adopt a similar approach but incorporate content, with
information from links and node attributes contributing equally to
module discovery.}
% We extend this approach by incorporating content. This way information from links and node attributes equally contribute to discovering modules.
Below, we briefly describe the Map Equation method.

\subsection{Compressing Random Walks}
We first need a method to encode the path traversed by a random walk on a network.  Consider the set of nodes, and assign to each node a codeword, such as a Huffman code~\cite{Huffman}. Huffman encoding gives more frequent codewords a shorter length, whereas less common codewords get longer description lengths.  The length of a codeword is taken to be the number of bits required to represent it. We expect the nodes that are visited more often by a random walk to have shorter codeword lengths.

Consider $X$, a random variable with $n$ possible states, where the $i$th state occurs with frequency $x_i$.  Then according to Shannon's source coding theorem~\cite{Shannon}, in order to describe the $n$ codewords representing the possible states, the average codeword length must be greater than or equal to the entropy
$$H(X) = -\sum_{i=1}^n x_i \log_2(x_i).$$
This is the basis for the Map Equation, which aims to minimize the full description length of a code based on the average codeword length.

The network description length is calculated at two levels, the node level and the module level.  A module is a group of nodes that have been merged, i.e., a community.  Without any modules, $n$ distinct codewords are required  to represent the $n$ nodes.  The more nodes, the longer the longest codeword will be.  Consider a partition of the nodes into $m$ modules.  For each module, there is a set of codewords to represent the nodes within the module, which can be reused in other modules. This shortens the length of the longest codeword that describes the nodes.

While the longest codeword for nodes is shorter, the description must now also take into account codewords to represent which module was entered by the path.  It may seem counterproductive to have two codewords to locate a single node.  However, when describing a path on a network, if the random walker remains in a particular module for a long time before switching modules, then the codewords for indicating module entrance are used less frequently.  Therefore, merging nodes into modules is advantageous when the nodes form a dense cluster with few links to other modules.  If this is the case, then the modules form communities where the information flow is greater within the module, highlighting the relationship of the nodes.

\subsection{The Map Equation}

The Map Equation~\cite{Rosvall08} gives the average description length for a step of an infinite random walk.   We now review the details for the Map Equation, a two-level description of the network.
At the first level, modules are connected to other modules.  At the second level, nodes are connected to others within the module.  Thus, we need to incorporate codewords at both levels into the Map Equation.

Given partition ${M}$ of $n$ nodes into $m$ modules, let  $\qiarrow$ be the probability of exiting module $i$.  Then  $\qarrow = \sum_{i=1}^m \qiarrow$ gives the probability that the random walk leaves a module in a given step.  From this, we find the entropy of the movement between the modules,
\begin{equation}
H(\Q) = -\sum_{i=1}^m \frac{\qiarrow}{\qarrow} \ \log_2\left( \frac{\qiarrow}{\qarrow} \right).
\end{equation}
This average codeword length is then weighted by the frequency with which a path exits a module, giving the first term in the Map Equation, $\qarrow H(\Q)$.

For the second term of the Map Equation, we look within each module and examine the possible steps for a random walker.    A random walker can either move to another node within the module or exit the module with probability $\qiarrow$.  Let $\palpha$ be the frequency with which node $\alpha$ is visited.  If we then consider the possible states for a random walker within module $i$, the movement entropy within the module is given by
\begin{eqnarray}
H(\Ptoi) =&-& \frac{\qiarrow}{\PplusQ} \ \log_2\left( \frac{\qiarrow}{\PplusQ} \right)\\
&-&\sum_{\alpha \in i}  \frac{\palpha}{\PplusQ} \ \log_2\left( \frac{\palpha}{\PplusQ} \right).
\end{eqnarray}
Each entropy term $H(\Ptoi)$ is then weighted by the frequency of being in one of these states, $$\piarrow  = \PplusQalpha.$$
The full Map Equation (ME) is given by
\begin{equation}\label{eq:MapEquation}
L(M) = \qarrow H(\Q) + \sum_{i=1}^m \piarrow H(\Ptoi).
\end{equation}
By minimizing this equation, the network description length is compressed while communities of nodes with higher information flow are identified.

%\section{Adding Node Attributes to In\-for\-mation Flow Description}
\section{Adding Node Attributes}
\label{sec:content}
%The network partition obtained through the graph compression technique of optimizing the Map Equation only uses the structure of the network to identify clusters of nodes.

The Map Equation uses only information in links to partition the network into modules. However, networks are often
\emph{content-rich},
meaning that nodes have attributes associated with them.  These attributes can provide more insight into the correct module classification of nodes, and they contribute to the description of information flow on a network. Take, for example, the world wide web. In addition to structure, given by hyperlinks between web pages, each page contains content, e.g., words, that differentiate it from other pages. Taking content into account gives a more robust view of the structure of the world wide web. Here we propose the Content Map Equation, which incorporates information about node attributes into the description of the random walk. This description can then be minimized to find modules in rich networks.

\subsection{The Content Map Equation}

%As described earlier, many of the existing community detection approaches incorporate content information by adding virtual edges into the network. Here, we consider an alternative way to add on the description length of content to Equation~(\ref{eq:MapEquation}).

We explicitly add the description length of node attributes into the Map Equation.
We first consider a dictionary for node $\alpha$, $\{\djalpha\}$, that consists of attributes associated with the node. We then create a dictionary vector $\xvalpha$  that gives the relative weight of each attribute for node $\alpha$, i.e., $\sum_{j}~\xalpha_j~=~1$. For example, when attributes are words from text associated with the node, the weight could simply be the frequency of each word. Next, we define a vector for each module, consisting of the dictionary vectors weighted by the node visit frequency, namely
\begin{eqnarray}
\xmodule_j = \sum_{\alpha\in i} \palpha ~ \xalpha_j.
\end{eqnarray}

We examine the possible content states for the random walker within a module.  The importance of attribute $j$ in module $i$ is given by
$\frac{\xmodule_j}{\pmodule},$
where $\pmodule = \sum_{\alpha\in i}\palpha$.  Thus, the average codeword length for the dictionary attributes within module $i$ is bounded below by the entropy,
\begin{eqnarray}
H(\Xtoi) &=& -\sum_{j} \frac{\xmodule_j}{\pmodule} ~ \log_2\left( \frac{\xmodule_j}{\pmodule}  \right).
\end{eqnarray}
This quantity is then weighted by the frequency of being in module $i$, $\pmodule$.

%With this \remove{description length of content information associated with nodes} description of a path,
We add the term above to Eq.~\ref{eq:MapEquation}, resulting in the Content Map Equation (CME):
\begin{equation}\label{equ:LCM}
%\begin{aligned}
L_C(M) =q_{\curvearrowright} H(\Q) + \sum_{i=1}^m p_{\circlearrowright}^{i} H(\Ptoi)  + \sum_{i=1}^m \pmodule H(\Xtoi).
%\end{aligned}
\end{equation}
This gives the average description length of a step of an infinite random walk on a network with node attributes.

Note that this method has several desirable properties. The foremost advantage of the approach is its simplicity. It does only one thing --- minimize the description length of a random walk --- to partition the network using information from both links and node attributes.
%The method prefers to group together nodes with the same attributes.
Furthermore, results do not depend on the number of attributes used to characterize nodes. This means that although a bad choice of representation (e.g., duplicating each attribute) would change the average description length, it will not affect partitioning results. Finally, since both links and attributes contribute equally to representing information in a network, our method does not require an additional parameter to control the contribution of each, in contrast to other methods~\cite{icde/QiAH12,wwwRuanFP13,BACG12,zhouCY09,getoor13}.

%==========================================================
\subsection{Illustrative Examples}\label{subsec:example}
%==============================================================
%We have discussed our extension of the Map Equation by adding content information, now let us use some illustrative examples to demonstrate more how the addition of the content description impacts the division of the network.

\begin{figure}[!htb]
\begin{center}
\begin{tabular}{@{}cc@{}}
\includegraphics[width=0.48\columnwidth]{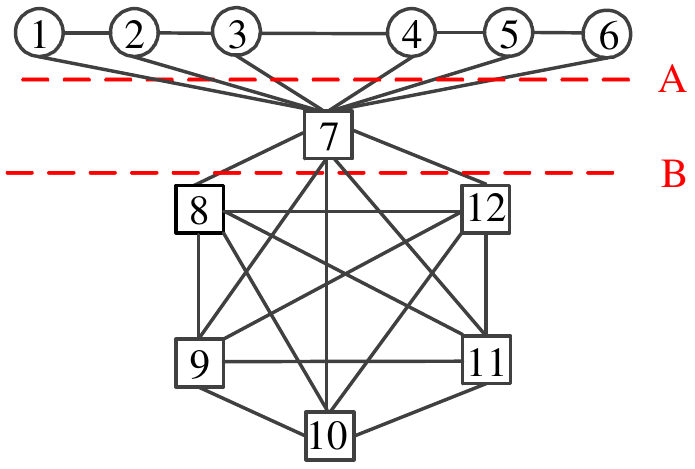} &
\includegraphics[width=0.48\columnwidth]{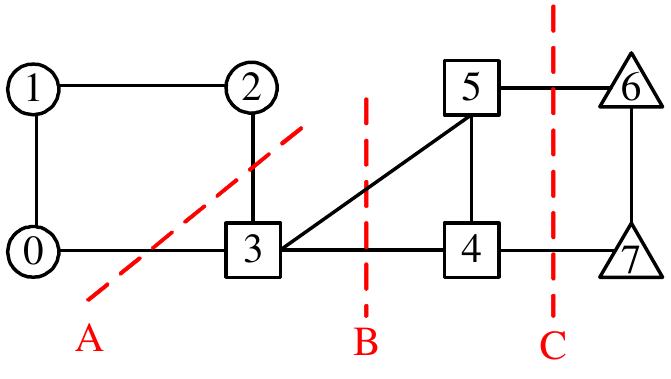} \\
(a) & (b)
\end{tabular}
\caption{Example networks containing  (a) two and (b) three different node types. Nodes of one type are described by a unique attribute vector. Dashed lines indicate possible partitions of the network.
%An illustrative example (A clique connected to a line) of the Content Map Equation. The dashed lines B and C indicate possible divisions that will optimize the map and Content Map Equations, and different shapes indicate different content vectors.
\label{fig:KDDToy}}
\end{center}
\end{figure}

We demonstrate how adding content  to the Map Equation can improve module division of a network with two illustrative examples.
The first example consists of a clique, with one clique node connected to a chain of nodes, as shown in Figure~\ref{fig:KDDToy}(a). Dashed lines represent possible partitions of the network. Cut $A$ bisects the network into two modules, grouping node $7$ with the rest of its clique, whereas cut $B$ groups it with the chain of nodes $1-6$. For simplicity, we use symbols to represent distinct nodes with $d$ attributes described by vectors:
\begin{equation}
\begin{aligned}
\bigcirc&=& (\underbrace{2/d,\cdots}_{d/2}, \underbrace{0, \cdots}_{d/2})\ \ \ \
\Box&=& (\underbrace{0, \cdots}_{d/2}, \underbrace{2/d,\cdots}_{d/2})
\end{aligned}
\end{equation}

\begin{table}[!hbt]
  \centering
\caption{Minimum average description lengths of  different partitions of the network in Figure~\protect\ref{fig:KDDToy}(a) using links along (ME), attributes alone, or both links and attributes (CME) in the description of information flow. The number of node attributes is $d=4$ or $d=1,000$.}\label{tab:MDLtoy2}
\begin{tabular}{|c|c|c|c|c|c|c|}
  \hline
%Cut&$L(M)$&$\sum_{i=1}^m\pmodule H(\Xtoi)$&$L_C(M)$\\
\emph{Cut}&\emph{ links} &\multicolumn{2}{c|}{\emph{attributes}} &\multicolumn{2}{c|}{\emph{both (CME)}}\\
\cline{3-6}
&\emph{(ME)}&$d=4$&$d=1000$&$d=4$&$d=1000$\\
\hline
no cut&3.41&1.89&10.86&5.30&14.27\\
\hline
A&3.59&\textbf{1.00}&\textbf{9.97}&\textbf{4.59}&\textbf{13.55}\\
\hline
B&\textbf{3.36}&1.51&10.47&4.87&13.83\\
\hline
\end{tabular}
\end{table}

Table~\ref{tab:MDLtoy2} gives the number of bits required to describe different partitions of the network using the Map Equation and Content Map Equation for different number of node attributes. Let us first consider the case with $d=4$ attributes and no cuts. It takes 3.41 bits to describe information contained in the links only using the Map Equation. On the other hand, it takes 1.89 bits to describe attributes alone, which indicates there is less information in the attributes than in the links.

Incorporating attributes changes the optimal partition of the network. Without attributes, the Map Equation prefers a balanced cut and chooses cut $B$ over cut $A$ or no cut at all, since it requires fewer bits (3.36). However, when attributes are incorporated into the Content Map Equation, cut $A$ has a lower description length (4.59 bits) than cut $B$ (4.87 bits). This partition is more consistent with our intuition for grouping node $7$ with similar nodes in the clique.

The Content Map Equation works correctly as the number of attributes grows. When there are $d=1000$ attributes, it takes 10.86 bits to encode content alone, compared to  3.41 bits to describe links. However, it still prefers cut $A$ over cut $B$.

Next we consider a more complex network in Figure~\ref{fig:KDDToy}(b) with three distinct node types, given by vectors:
\begin{equation}
\begin{aligned}
 \bigcirc&=& (\underbrace{2/d,\cdots}_{d/2}, \underbrace{0, \cdots}_{d/2})\\
\Box&=&(\underbrace{0, \cdots}_{d/2}, \underbrace{2/d,\cdots}_{d/2})\\
 \triangle&=&(\underbrace{0, \cdots}_{3d/4}, \underbrace{4/d,\cdots}_{d/4})
 \end{aligned}
\end{equation}

\noindent Note that circles and squares do not share any attributes, while squares and triangles share some attributes.

\remove{
\begin{figure}[!htb]
\begin{center}
\includegraphics[width=0.4\textwidth]{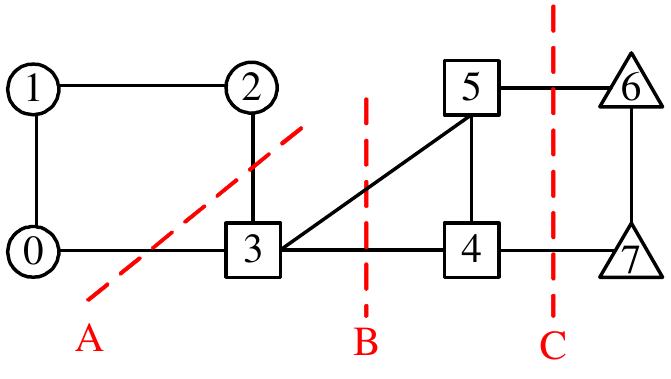}
\caption{Another illustrative example of the Content Map Equation. The dashed lines A, B, and C indicate possible divisions that will optimize the map and Content Map Equations, and different shapes indicate different content vectors. \label{fig:contenttoy1}}
\end{center}
\end{figure}
}

\begin{table}
\centering
\caption{Minimum average description lengths of  different partitions of the network in Fig.~\protect\ref{fig:KDDToy}(b), using links along (ME), attributes alone, or both links and attributes (CME) in the description of information flow. The number of node attributes is $d=4$ or $d=1,000$.}\label{tab:MDLtoy1}
%\scalebox{0.9}
{
\begin{tabular}{|c|c|c|c|c|c|c@{}|}
  \hline
%Cut&$L(M)$&$\sum_{i=1}^m\pmodule H(\Xtoi)$&$L_C(M)$\\
\emph{Cut}&\emph{ links} &\multicolumn{2}{c|}{\emph{attributes}} &\multicolumn{2}{c|}{\emph{both (CME)}}\\
\cline{3-6}
&\emph{(ME)}&$d=4$&$d=1000$&$d=4$&$d=1000$\\
\hline
no cut&2.95&1.84&9.81&4.79&12.75\\
\hline
A&3.02&0.96&8.92&\textbf{3.98}&\textbf{11.95}\\
\hline
B&\textbf{2.93}&1.43&9.39&4.35&12.32\\
\hline
C&3.15&1.56&9.53&4.72&12.68\\
\hline
A+C&3.27&\textbf{0.80}&\textbf{8.77}&4.07&12.03\\
\hline
A+B&3.18&0.94&8.91&4.12&12.08\\
\hline
B+C&3.21&1.29&9.25&4.50&12.46\\
\hline
A+B+C&3.61&\textbf{0.80}&\textbf{8.77}&4.41&12.38\\
\hline
\end{tabular}
}
\end{table}

Table~\ref{tab:MDLtoy1} gives the minimum average number of bits required to describe information in this network when taking into account only links, only content, and both links and content. Without the attributes, the Map Equation (ME) chooses cut $B$ over the other options, partitioning this network into two modules. This seems like the natural, balanced division of the network. However, including content information changes the preferred partition of the network. While there are three distinct vectors, the nodes 3-7 share some of the attributes. Thus, when minimizing the Content Map Equation (CME) we find that the best solution is cut $A$ with 3.98 bits when there are $d=4$ attributes. This is different from either the cut preferred by links alone or the cut preferred by attributes alone. When nodes have many ($d=1000$) attributes, cut $A$ with 11.95 bits is still the preferred cut.
Thus, partitioning results do not depend on the number of attributes used to characterize nodes.

\remove{Alternatively, if the dictionaries of squares and triangles also do not share any features,
{\color{blue}\begin{eqnarray}
 \bigcirc&=& \left( 0.5, 0.5, 0, 0,0,0\right)\\
\Box&=& \left( 0, 0, 0.5, 0.5,0,0 \right)\\
 \triangle&=& \left( 0, 0, 0, 0,0.5,0.5 \right).
\end{eqnarray}}
then the optimal solution is cut $A+C$ with minimum 4.27 bits. In addition, we notice that we only needs 0.8 bits to encode the content when squares and triangles have overlapping but 1.0 bits to encode the content when they do not have overlapping. This is because we are encoding the difference of content information distribution, but not the content information itself.}

\section{Finding Modules}\label{sec:greedy}
% Optimization problem
Minimizing the Content Map Equation is an NP-hard optimization problem. Similar to minimizing the matrix storage cost~\cite{SDM12}, the difficulty can be established by reducing it to the traveling salesman problem. To this end, we study feasible solutions from the realm of iterative heuristic algorithms. Rosvall \& Bergstrom used an agglomerative (bottom-up) method that begins with each node in its own module and proceeds by greedily merging modules so as to decrease the description length. Unfortunately, even this greedy method is too computationally complex for larger networks. To address this issues, we further propose a scalable solution, namely top-down greedy search (see Section~\ref{subsec:topdown}).

\subsection{Bottom-up Method}
\label{subsec:bottomup}
We first consider a greedy agglomerative, or bottom-up, search algorithm~\cite{Rosvall08}, where each node is initially placed in its own module. Then, at each iteration, we merge two modules that result in the largest decrease in the Content Map Equation.  This is repeated until there is no further benefit to merging modules. The details are presented in Algorithm~\ref{alg:bottomup}.

\begin{algorithm}[!t]
\caption{Bottom-up search for Content Map Equation}\label{alg:bottomup}
\begin{tabbing}
\textbf{Input}: Network $G$\\
\textbf{Output}: A partition $M$ of $G$\\
~1: \textbf{for} each node $\alpha$\\
~2: \hspace{0.5cm}compute $\palpha$ and $\xvalpha$\\
~3: Initialize $M$ by assigning each node to its own module\\
~4: $\Delta L=0$\\
~5: \textbf{do}\\
~6: \hspace{0.5cm}let $M_{ij}$ denote a new partition resulting from merging \\
\hspace{1.05cm}modules $i$ and $j$ from $M$\\
~7: \hspace{0.5cm}\{$x$,$y$\}=$\arg\min\limits_{\mbox{$i$, $j$}}$ ($L_C(M_{ij})$-$L_C(M)$)\\
~8: \hspace{0.5cm}$\Delta L$=$L_C(M_{xy})$-$L_C(M)$\\
~9: \hspace{0.5cm}set $M$ as $M_{xy}$ if $\Delta L<$0\\
10: \textbf{while} $\Delta L<0$\\
11: \textbf{return} $M$
\end{tabbing}
\vspace{-4mm}
\end{algorithm}

In lines 1--2, we first calculate $\palpha$ and $\xvalpha$ for each node. These quantities are constant and independent of the partition. The vector $\xvalpha$ is chosen to give the weight of each attribute (or frequency of a word) associated with node $\alpha$.  If common attributes are shared by many nodes, then it may be more appropriate to use tf-idf weighing, lessening the importance of attributes associated with multiple classes.

The steady state of the node visit frequency of the infinite random walk, $\palpha$, can be easily approximated for directed networks with the PageRank algorithm~\cite{PageRank}.  A small probability of teleportation to random nodes can be introduced to guarantee a unique steady state.  Rosvall \& Bergstrom~\cite{Rosvall08} chose $\tau=0.15$, which is equivalent to a damping factor of $0.85$.  For undirected networks, this node visit frequency is the relative sum of the edge weights incident to node $\alpha$, compared to twice the full edge weight of the network, namely
\begin{equation}
\palpha = \frac{\sum_{\beta=1}^n A_{\alpha, \beta}}{\sum_{\beta=1}^n\sum_{\gamma=1}^nA_{\beta, \gamma}},
\end{equation}
where $A$ is the weighted adjacency matrix of the undirected network, with values corresponding to the edge weights between incident nodes.

After initialization, we start the greedy search (lines 5--10). The critical part is to compute the $L_C(M)$ (Eq.~\ref{equ:LCM}) for each possible partition $M_{ij}$, especially the exit probabilities for a given step $\qiarrow$, which can be easily calculated by
\begin{equation}
\qiarrow = \tau \left(\frac{n-n_i}{n-1}\right)\sum_{\alpha \in i} \palpha + (1-\tau) \sum_{\alpha \in i}\sum_{\beta \notin i} \palpha A_{\alpha, \beta}
\end{equation}
for directed networks and
\begin{equation}
\qiarrow =\sum_{\alpha \in i}\sum_{\beta \notin i} \palpha A_{\alpha, \beta}
\end{equation}
for undirected networks.  Here, we take $A$ to have row sums of one.

While this method does not provide the optimal solution to the minimization problem, it gets a reasonable approximation that identifies clusters of nodes with similar attributes as well as local structures.

\subsubsection{Convergence analysis}
We now briefly analyze the convergence property of the bottom-up algorithm. The Content Map Equation has both lower and upper bounds. In addition, the total cost of Eq.~\ref{equ:LCM} is monotonically decreasing using Algorithm~\ref{alg:bottomup}, since two modules are merged if and only if the total cost can be reduced, and the stopping criterion is satisfied if and only if the total cost cannot be reduced any further. Thus, the bottom up algorithm converges to a local optimum.

\subsubsection{Complexity analysis}
The computational complexity of each iteration of the bottom-up algorithm is $O(m^2(n+l+d))$, where $m$ is the number of modules, $n$ is number of nodes, $l$ is number of links, and $d$ is number of attributes. Hence, the total complexity of the partitioning procedure is $O(km^2(n+l+d))$, where $k$ is number of iterations, which is usually a small number. Note that in the bottom-up algorithm, we start from the state where each node is a module, that is, in the worse case, $m=O(n)$.

\subsection{Top-down Method}
\label{subsec:topdown}
In the bottom-up method, we compute a better partition $M$ with $m$ modules from a partition $M^{\prime}$ with $m+1$ modules. However, for the initial state $m=n$, and the search space is essentially quadratic in the network size. For networks with a large number of nodes, the computational costs of even the greedy algorithm may be prohibitive. To address this problem, we propose a ``top-down'' search algorithm.
%We also discuss the advantage of the top-down approach over the bottom-up approach in large graphs for computational complexity.

At first glance, it may be preferable to start with all nodes in the same module and proceed by splitting modules until no further decrease in the description length is achieved. However, in reality, this method can easily get trapped in local minima that do not represent a good partition of the network. Instead, we start from a random configuration, with nodes assigned randomly to $m=\sqrt{n}$ modules. Note that this random configuration does not mean the number of modules found by the algorithm is $\sqrt{n}$ since both splitting (when a node is moved to a new empty module) and merging (all of the nodes in a module are assigned to another module) are considered in the algorithm. In each iteration of the search algorithm, a node is either assigned to a different existing module or a new empty module, whichever leads to a larger decrease in CME. The algorithm stops when it can no longer decrease the description length.

\begin{algorithm}[!t]
\caption{Top-down search for Content Map Equation}\label{alg:topdown}
\begin{tabbing}
\textbf{Input}: Network $G$\\
\textbf{Output}: A partition $M$ of $G$\\
~1: Initialize $\Delta L=0$, $M$,  $\palpha$ and $\xvalpha$\\
~2: \textbf{for} $i=1$ to $\sqrt{n}$\\
~3: \hspace{0.5cm}randomly initialize partition $M^{\prime}$ with $\sqrt{n}$ modules\\
~4: \hspace{0.5cm}set $M$ as $M^{\prime}$ if reducing description length\\
~5: \textbf{do}\\
~6: \hspace{0.5cm}\textbf{for} each $\alpha$ in ordered node list $V$\\
~7: \hspace{1.0cm}let $M(\alpha)_{i}$ denote the new partition resulting\\
\hspace{1.55cm}from moving node $\alpha$ to an existing or a new \\
\hspace{1.55cm}empty module $i$\\
~8: \hspace{1.0cm}$x$=$\arg\min\limits_{i}$ ($L_C(M(\alpha)_{i})$-$L_C(M)$)\\
~9: \hspace{1.0cm}$\Delta L=L_C(M(\alpha)_{x})-L_C(M)$\\
10: \hspace{1.0cm}set $M$ as $M(\alpha)_{x}$ if $\Delta L<0$\\
11: \textbf{while} $\Delta L<0$\\
12: \textbf{return} $M$
\end{tabbing}
\vspace{-4mm}
\end{algorithm}

The top-down search is detailed in Algorithm~\ref{alg:topdown}. In lines 2--5, we create a random partition and choose the one with the smallest description length as the start state. While this heuristic is simple and na\"{\i}ve, it achieves better performance in real data than using LDA~\cite{BleiNJ03} or ME as initializations (see Section~\ref{subsec:Init}). Next, for each node $\alpha$, we enumerate all possible improvements for $\alpha$: assign $\alpha$ to either another existing module or a new empty module (lines 7--11). Another heuristic strategy in our algorithm is that we notice the previous correction for a node $\alpha$ might have influence for the latter correction of another node $\beta$. Hence, in each iteration, we order the node lists based on the descending order of $\palpha$ (line 8). The high-level intuition is that we want to find the improvements for those highly influential nodes first and then turn our attention to less influential nodes.
The top-down search algorithm is guaranteed to converge to a local optimum. Convergence properties are similar to the bottom-up algorithm.

\subsubsection{Complexity analysis}
In each iteration, for each node $\alpha$ and each module $i$, we need to compute the change in description length, $L_C(M(\alpha)_{i})$ -$L_C(M)$. This computation is very efficient, since only the source and target module is affected, thus the time complexity is $O(\delta_n+\delta_l+\delta_d)$. Here $\delta_n$, $\delta_l$, and $\delta_d$ denote the average number of nodes, links, and attributes respectively in the neighborhood of a node. Then the overall time complexity of our algorithm is $O(knm \times (\delta_n+\delta_l+\delta_d))$, where $k$ is the number of iterations. In practice, however, both $m$ and $k$ are very small. Thus, our partitioning algorithm is efficient in most cases, as also verified by our experiments.

%=============================================================================
\section{Evaluation on Real-World Data}
\label{sec:results}
%============================================================================
We use the Content Map Equation to partition real-world networks with known ground truth community labels. All of these networks are examples of content-rich networks in which nodes have attributes, such as content words for scientific papers in citation networks, or demographic features for Facebook users.

\subsection{Data Sets}
\textbf{Twitter:} We consider a network created by interactions among Twitter users on the subject related to proposition 30 on the November 2012 California ballot~\cite{Smith13socialcom}. We used the method described in \cite{Smith13socialcom} to classify the position on the proposition of each user as for, against or neutral. These serve as the ground truth labels for these data.

For the attributes associated with each node (user), we considered the 25 hashtags used most frequently by that user and used tf-idf scores instead of hashtag frequency in the nodes' attribute vectors.

\textbf{Facebook:} We used a subset of a large social network of Facebook users that contains anonymized information about individuals, including hometown, gender, major, work, and year in school~\cite{Facebook}.  We took these features as attributes of each node.  An edge represents a friendship between two users.  For the ground truth community labels, we used the circles that have been identified in these data~\cite{Facebook}, with some users being members of multiple circles and other users not in any circle.

\textbf{ArnetMiner:}
The ArnetMiner dataset is a citation network~\cite{ArnetMiner}, classified according to research fields: data mining and association rules (DM), database systems and XML data (DB), information retrieval (IR), web services (WS), bayesian networks and belief function (BN), web mining and information fusion (WM), semantic web and description logics (SW), machine learning (ML), pattern recognition and image analysis (PR), natural language system and statistical machine translation (NLP). We used these class labels as ground truth data in the experiment. We treated words in the paper title as node attributes.
%The most frequently observed combination is topics 24 and 107, occurring 17 times.

%For any citation network, there are several possible dictionaries that can be used, including titles, abstracts, author names, journal of publication, keywords, categories and subject descriptions, and general terms.
%Here, we take the simplest of these, the titles, creating a dictionary consisting of  4,214 words.  For each individual paper, its dictionary is comprised of words from its title.

\textbf{Citeseer:}
The CiteSeer dataset~\cite{sen:aimag08} is a citation network with 3312 scientific papers, classified into one of six classes, and 4732 links. Each paper is described by a 0/1-valued vector indicating the absence/presence of the corresponding word from the dictionary. The dictionary consists of 3703 unique words.

\textbf{Pubmed:} The Pubmed Diabetes dataset~\cite{sen:aimag08} contains 19,717 scientific publications from the PubMed database pertaining to diabetes, classified into one of three classes. Each publication in the dataset is described by a tf-idf weighted word vector from a dictionary which consists of 496 unique words.

\textbf{Flickr:}
This dataset~\cite{conf/eccv/McAuleyL12} was built by creating links between images from Flickr that share common metadata: images from the same location, submitted to the same gallery, group, or set, images taken by friends, etc.  The attributes of a single node (image) include image features that are obtained from PASCAL~\cite{pascal-voc-2011}, ImageCLEF~\cite{Imageclef10}, MIR~\cite{huiskes08}, and NUS-wide\cite{nus-wide-civr09}. We use the ground truth labels in the image classification tasks as the ground truth communities (only around 10\% nodes have ground truth communities).

A set of selected statistics of all the above data are reported in Table~\ref{tab:datastat}.

\begin{table}[t]
  \centering
  \small{
  \caption{Statistics of datasets.}\label{tab:datastat}
\begin{tabular}{|c|c|c|c|c|}
\hline
&\#nodes&\# links& \# classes& \# attributes\\
\hline
Twitter&565&1,008&3&24\\
\hline
ArnetMiner&2,555&6,101&10&4,214\\
\hline
Citeseer&3,312&4,536&6&3,703\\
\hline
Facebook&1,911&24,975&9&570\\
\hline
Pubmed&19,717&44,338&3&496\\
\hline
Flickr&105,938	&2,316,948&215&26,041\\
\hline
\end{tabular}
 }
\end{table}

\subsection{Qualitative Evaluation}
\begin{figure}[!bt]
\centering
\begin{tabular}{cc}
\includegraphics[width=0.51\columnwidth]{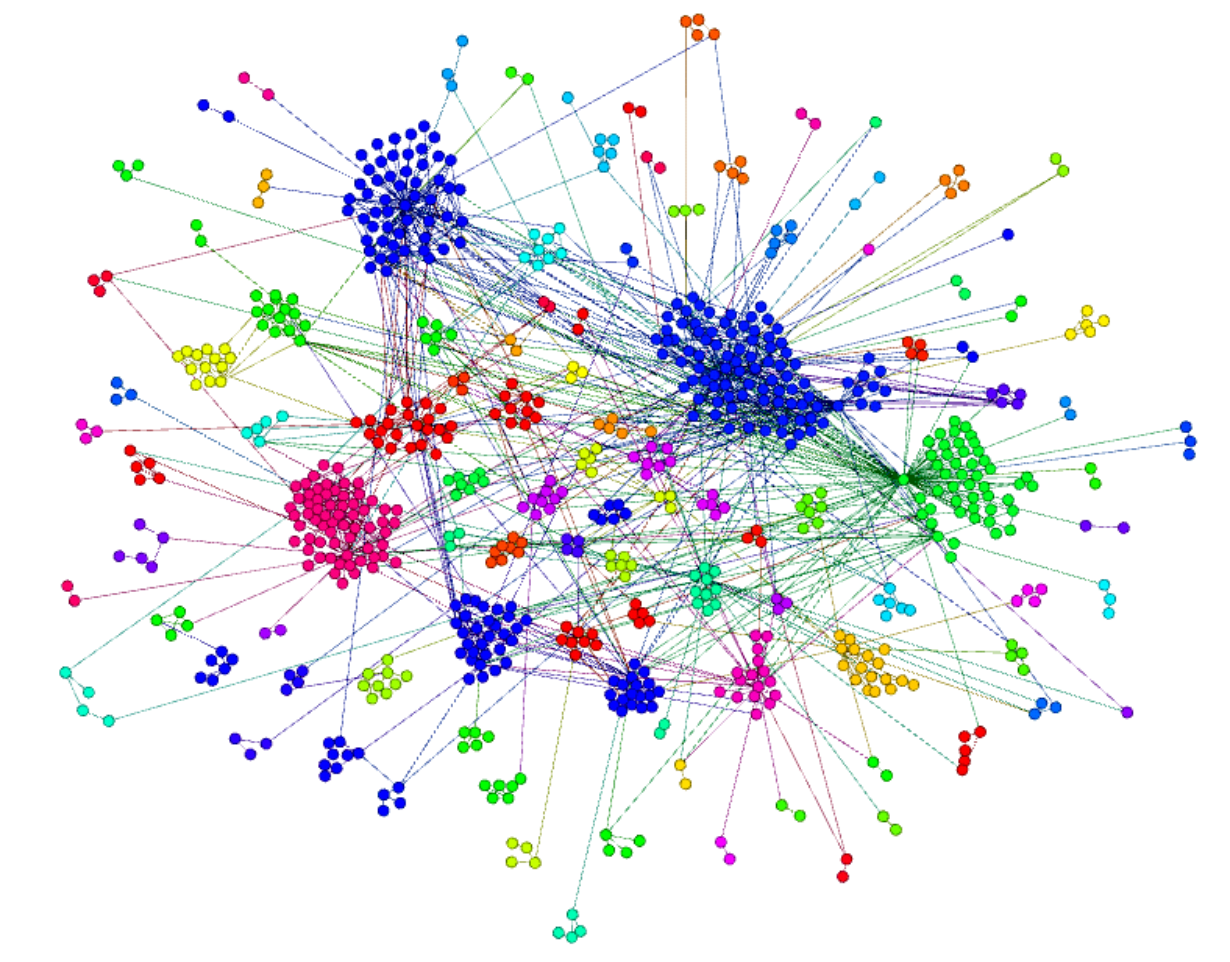} &
\hspace{-0.22cm}
\includegraphics[width=0.51\columnwidth]{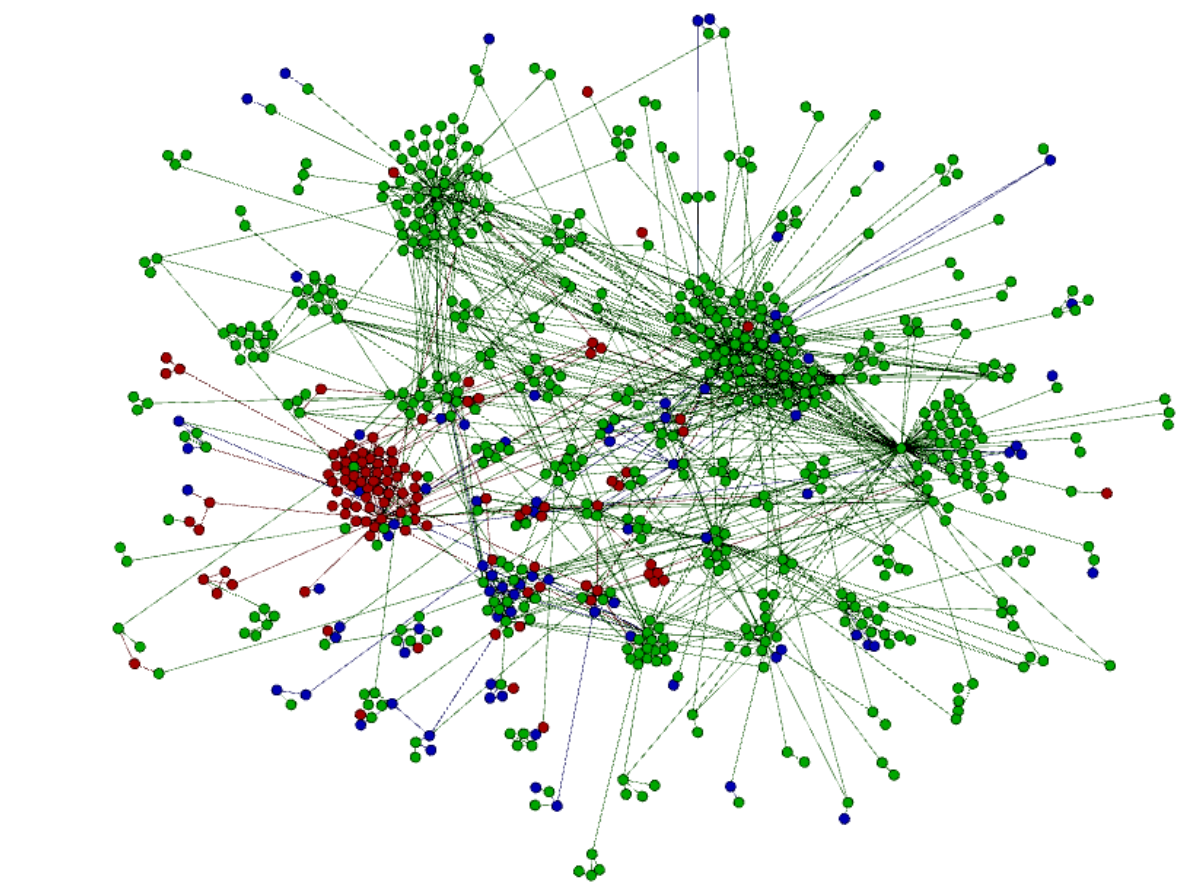}
\\
(a) modules & (b) ground truth
\end{tabular}
\caption{Largest component in the Proposition 30 user network. (a) Communities, indicated by different colors and locations,  found by the Content Map Equation. (b) Ground truth labels of the users in the network in (a): for (green), against (red), neutral (blue)(best viewed in color.)}\label{fig:Yes30}
\end{figure}

We first look at the Twitter interaction network for Proposition 30. Figure~\ref{fig:Yes30}(a) divides the network into communities according to the Content Map Equation, placing nodes within the same module closer together and with the same color. There are many small communities and a few larger, densely connected communities. But how far are the outputted communities from the ground truth?

The ground truth for this network is shown in Figure~\ref{fig:Yes30}(b), where we placed nodes in the same locations, but colored them according to their stance: green for users who support Proposition 30, red for users who oppose it, and blue for neutral users. This highlights the types of communities found in this network, including a few large communities that predominantly consist of users of one stance and a few smaller communities comprised of individuals with difference stances.  While the CME breaks users into many communities, the communities themselves are relatively pure, i.e., composed of users who have the same stance on the proposition.
%Of the latter, the largest community of 30 nodes is comprised of 24 individuals associated with news media, some of which have an observable opinion but tend to post more neutral stories.  Upon further inspection of the tweets by the other 6 individuals, a strong sentiment on Proposition 30 is not observed with the majority of the posts consisting of retweets or direct mentions of news media accounts.   Thus, this community obtained by the content map equation was able to identify a cluster of news media user accounts that tend to tweet about both sides of the proposition.

\begin{figure}[htb]
\centering
\includegraphics[ width=0.9\columnwidth]{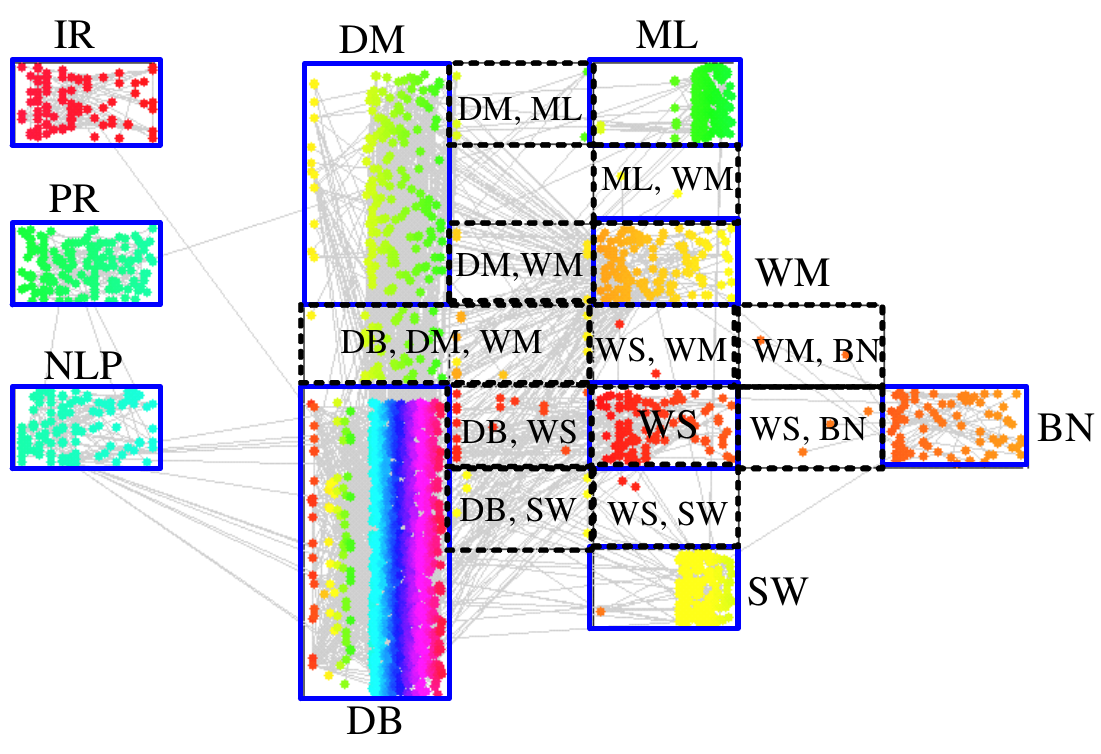}
\caption{ ArnetMiner citation network. Blue squares indicate different topics, with nodes inside the squares identified solely with that topic. Nodes inside dashed squares are associated with multiple topics and are placed between both topics with a hyphenated label. Color and location differentiate nodes in different communities found by the CME (best viewed in color).}\label{fig:ArnetMinerFull}
\end{figure}

Next, we visualize the communities found by CME in the ArnetMiner citations network. From Figure~\ref{fig:ArnetMinerFull}, we are able to observe that the communities are of two types, either a community with nodes of only one topic or a community with a mixture of topics. For instance, the majority of orange color nodes corresponds to the topic web mining and information fusion (WM). In communities of the latter kind, for instance, black dashed box with red color, it contains not only single-topic nodes of both (WS) and (DB), but also nodes that were members of both topics. One of the reason is that in the ground truth, some of the topics co-occur very frequently (i.e., there exists a single node with multiple ground truth topic labels). Thus, by qualitative evaluation, we verify that the partitioning outputted by CME correctly identifies similar nodes.

\subsection{Quantitative Evaluation}

%\subsection{Baselines and Evaluation Metrics}
We quantitatively evaluate network partitioning using the point-wise normalized F-measure, purity and clustering accuracy to compare how well the discovered communities reproduce the classes present in the data.
\paragraph{F-measure} Given an output community $p$ and with reference to a ground truth class $g$ (both in the form of node set), we define the precision rate as $|p\cap g|/|p|$ and the recall rate as $|p\cap g|/|g|$. The F-measure of $p$ on $g$, denoted as $F(p, g)$, is the harmonic mean of precision and recall rates. The final F-measure~\cite{wwwRuanFP13} of the outputted partitioning $P$ on the ground truth clustering $G$ is then calculated as
\begin{equation}\label{equ:fmeasure}
F(P,G)=\sum_{p\in P}\Big\{\frac{|p|}{n}\max_{g\in G}F(p,g)\Big\}.
\end{equation}

\paragraph{Purity} The purity of the outputted partitioning $P$ on the ground truth clustering $G$ is defined as
\begin{equation}\label{equ:purity}
\texttt{purity}(P,G)=\texttt{avg}_{p\in P}\Big\{\max_{g\in G}\frac{|p\cap g|}{|p|}\Big\}.
\end{equation}

\paragraph{Clustering accuracy}Assume that we assign the outputted community with ground truth label using the majority vote.  Then the clustering accuracy evaluates the percentage of nodes with correct assignments.
\begin{equation}\label{equ:accuracy}
A(P,G)=\frac{1}{n}\sum_{p\in P}\max_{g\in G}|p\cap g|
\end{equation}
We do not consider Normalized Mutual Information as a performance measure, because our approach finds a much larger number of classes than exist in the ground truth, making discovered classes poor predictors of the ground truth class distribution.
If there are two groups with the same ground truth label but with no edges between them, then we shouldn't expect them to be placed in the same module.

\paragraph{Baselines} We compare the algorithms proposed in this paper, bottom-up Content Map Equation (B-CME) and top-down CME (T-CME) to three classes of baselines: 1) content-based approaches, such as topic modeling (e.g., LDA~\cite{BleiNJ03}); 2) structure-based approaches such as the Map Equation (ME~\cite{Rosvall08}); and 3) methods which use both links and attributes, such as BACG~\cite{BACG12} and Codicil~\cite{wwwRuanFP13}. We do not compare to approaches~\cite{CESNAICDM13,ZhuNC11} since they only support a single-attribute per node. Our method produces much better F-measure scores than \cite{getoor13} on two benchmark dataset Citeseer and Pubmed. However, we do not include it in the results since they did not apply it to other datasets. The experiments were performed on a 2.7GHZ Intel i-7 CPU with 8G of memory.

\subsubsection{Runtime}
%=========================Figure Running time======================//
\begin{figure}[htp]
\centering
  \includegraphics[width=0.9\columnwidth]{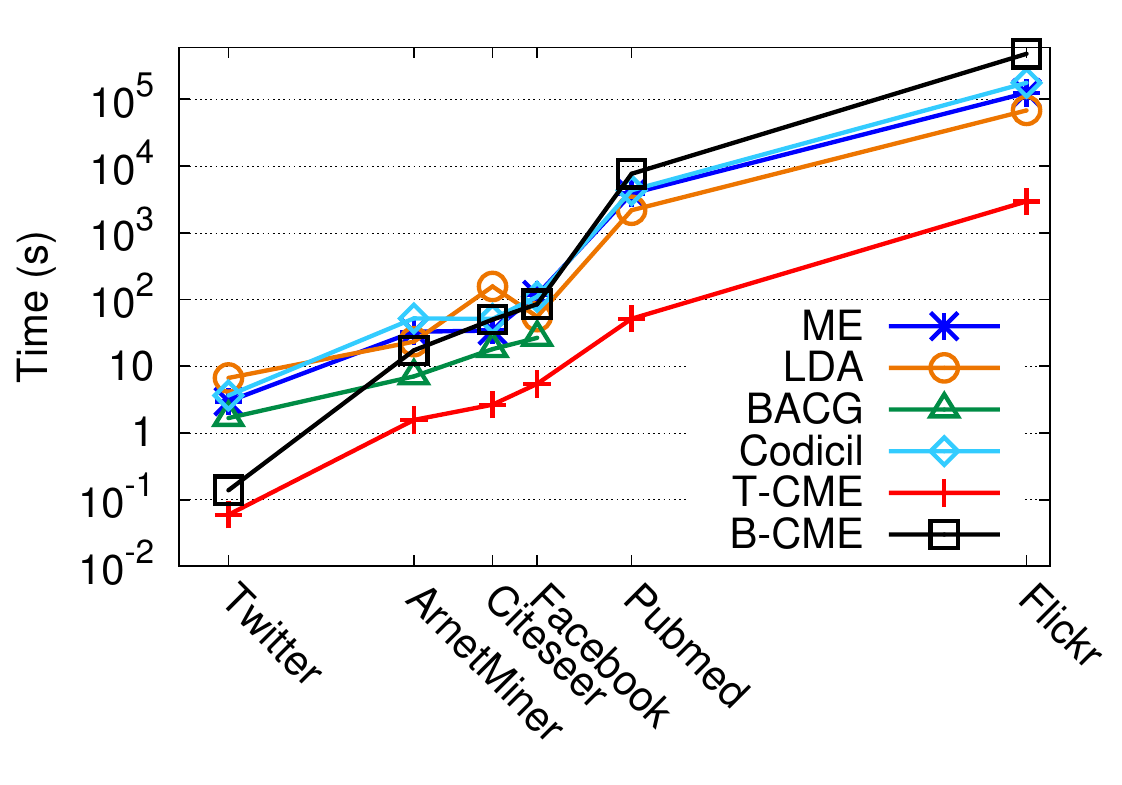}
  \caption{Runtime comparison of different algorithms. Data sets are ordered by their size (best viewed in color).}\label{fig:time}
  \end{figure}
  %=========================Figure Running time======================//
%We first evaluate the scalability of the proposed methods.
Figure~\ref{fig:time} compares the runtimes of different methods. Results are ordered by network size (number of nodes and links, see Table~\ref{tab:datastat}), except for the Citeseer citations network, which we put between ArnetMiner and Facebook datasets to improve visualization.

Note that for baseline BACG, we only have the results for the small to medium-size networks, since the implementation of BACG runs out of memory for large networks such as Pubmed and Flickr. Results indicate that our bottom-up search implementation is faster than other baselines for small networks, comparable to other baselines for medium size network, but is much slower than other baselines for large networks. This motivates us to use the top-down implementation, which is significantly faster than alternative methods. The T-CME is about one order of magnitude faster than other baselines and two orders of magnitude faster than the B-CME.

For the baselines, BACG is more efficient than others since it stores many matrices in memory to facilitate computation, which leads to its memory bottleneck for large networks. The running time of content-based approach LDA depends on the number of nodes and the number of attributes in content vectors. Hence, LDA runs much faster than others in Flickr and Facebook, where the link information is much heavier than the content information. Codicil first constructs a content graph, performs local sparsity analysis on the content graph, and then runs the community detection algorithm ME on the sparse content graph. Thus, Codicil always runs slower than ME due to additional costs to construct and sparsify the content graph.

\subsubsection{Performance}
%=========================Figure F measure======================//
\begin{figure}[!htp]
  \includegraphics[width=0.9\columnwidth]{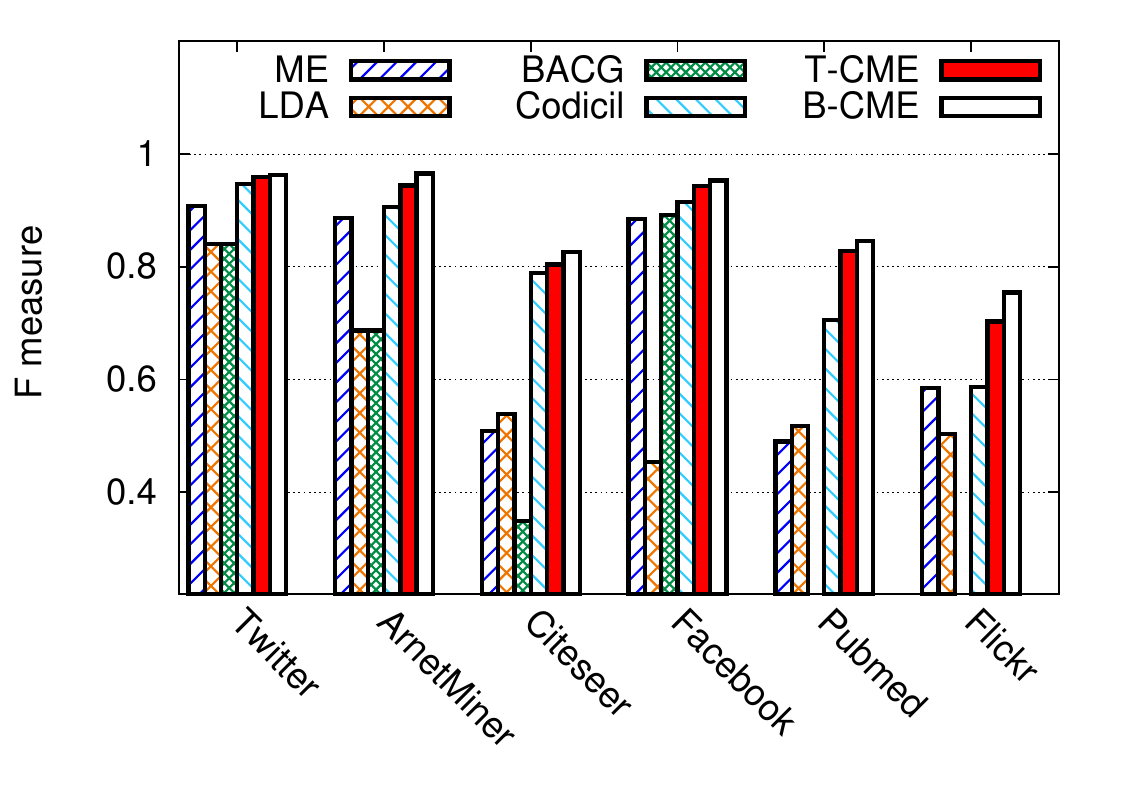}
  \caption{Performance comparison in terms of F measure (best viewed in color).}\label{fig:Fmeasure}
\end{figure}
%=========================Figure F measure======================//
Figure~\ref{fig:Fmeasure} compares the F measures (Eq.~\ref{equ:fmeasure}) obtained by different approaches on the datasets. Recall that BACG fails to run the Pubmed and Flickr datasets on our machine due to huge memory consumption.

The results indicate that inclusion of node attributes lead to a better partition than using links alone (ME). The improvement is especially dramatic for the Citeseer and Pubmed datasets. The possible reason is that in the Citeseer and Pubmed citations networks, each node has very few links on average; therefore, structural information is very weak. Hence, the Map Equation finds a worse grouping of papers than content-aware approaches. The Content Map Equation is also much better than using content alone (e.g., LDA).

Compared to baselines BACG and Codicil, the Content Map Equation (both top-down and bottom-up) is consistently better.  The top-down algorithm (T-CME) in general produces slightly worse results than the bottom-up approach, but it is much faster than the bottom-up method (B-CME). This indicates that the proposed top-down algorithm has a good trade-off between efficiency and quality.
 %=========================Figure Purity======================//
 \begin{figure}[!t]
 \centering
  \includegraphics[width=0.9\columnwidth]{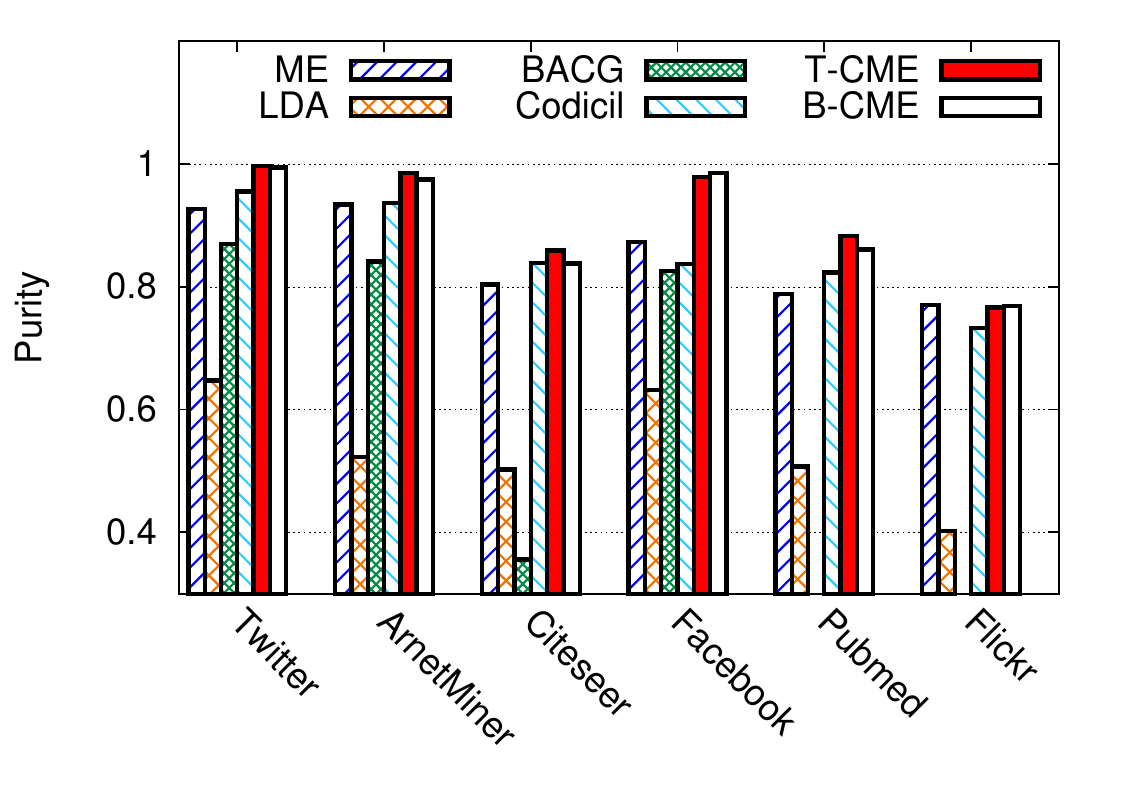}
  \caption{Performance comparison in terms of Purity (best viewed in color).}\label{fig:purity}
\end{figure}
%=========================Figure Purity======================//

%=========================Figure Accuracy======================//
\begin{figure}[!t]
\centering
  \includegraphics[width=0.9\columnwidth]{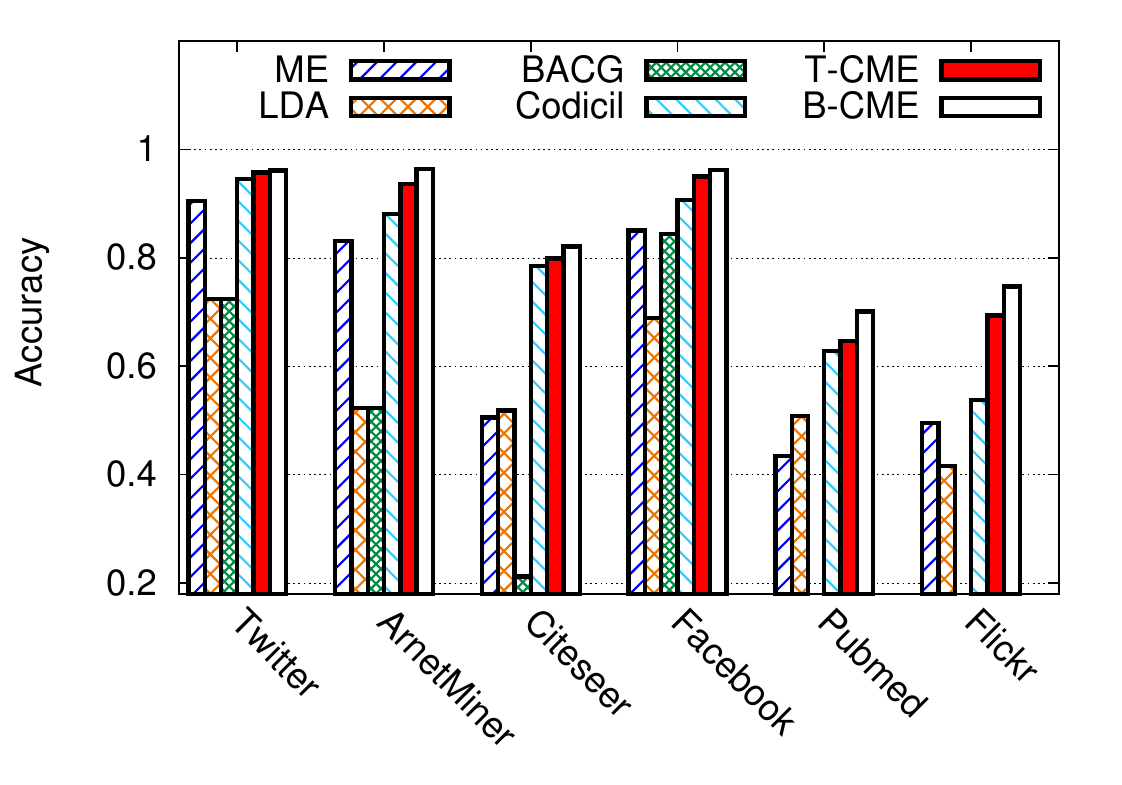}
  \caption{Performance comparison in terms of Accuracy (best viewed in color).}\label{fig:accy}
\end{figure}
%=========================Figure Accuracy======================//

\begin{table}[t]
  \centering
  \small{
  \caption{Average Minimum Description length (MDL).}\label{tab:MDLlength}
\begin{tabular}{|p{0.4in}|p{0.3in}|p{0.3in}|p{0.3in}|p{0.3in}|p{0.3in}|p{0.35in}|}
\hline
MDL&Twit ter&Arnet Miner&Cite seer&Face book&Pub med&Flickr\\
\hline
B-CME&7.780&12.354&12.033&13.524&15.870&15.79\\
\hline
T-CME&8.1436&12.596&11.601&14.120&15.809&16.092\\
\hline
\end{tabular}
 }
\end{table}

In addition to F-measure, we also use purity (Eq.~\ref{equ:purity}) and accuracy (Eq.~\ref{equ:accuracy}), shown in Figure~\ref{fig:purity} and Figure~\ref{fig:accy}, respectively, to evaluate network parittioning. Both results show that Content Map Equation outperforms the baseline BACG and Codicil. However, in Citeseer and Pubmed, the top-down greedy search (T-CME) produces a better partition than the bottom-up search (B-CME), according to the purity measure. There is no surprise if we look at the description length of the partitioning outputted by the two different search strategies (see Table~\ref{tab:MDLlength}). We notice that T-CME achieves lower description length than B-CME for Citeseer and Pubmed as well.  These results are consistent with the intuition~\cite{snobecluster} that if we can correctly categorize the data (high purity within cluster), then the data can be described with the highest efficiency (i.e., using the minimum message length).

In summary, our approach identifies better communities in content-rich networks than alternative state-of-the-art methods that also take links and node attributes into account.

\subsubsection{Optimizations: Effect of Initialization}\label{subsec:Init}
%=========================Figure Initialization results:F measure ======================//
\begin{figure}[htp]
\centering
\includegraphics[width=0.8\columnwidth]{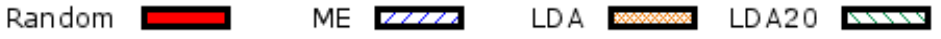}
\subfigure[]{\label{fig:InitFmeasure}\includegraphics[width=0.5\columnwidth]{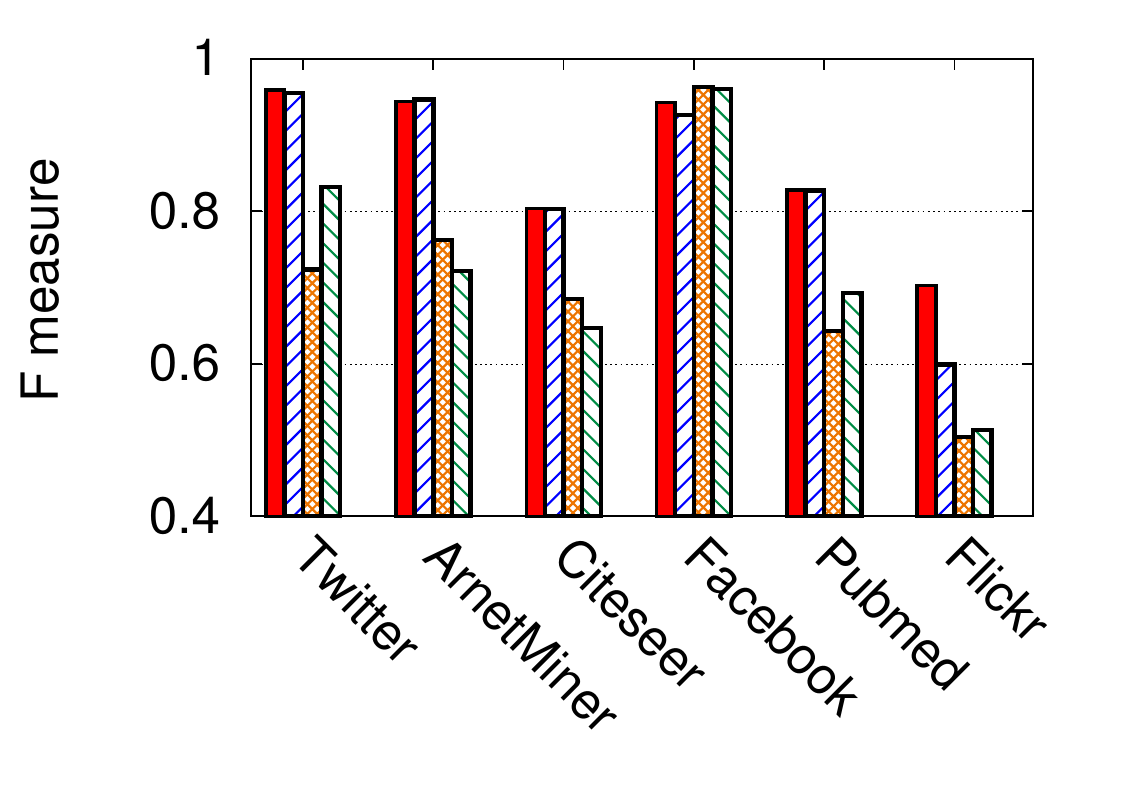}}
\hspace{-0.2cm}
\subfigure[]{\label{fig:Mlength}\includegraphics[width=0.5\columnwidth]{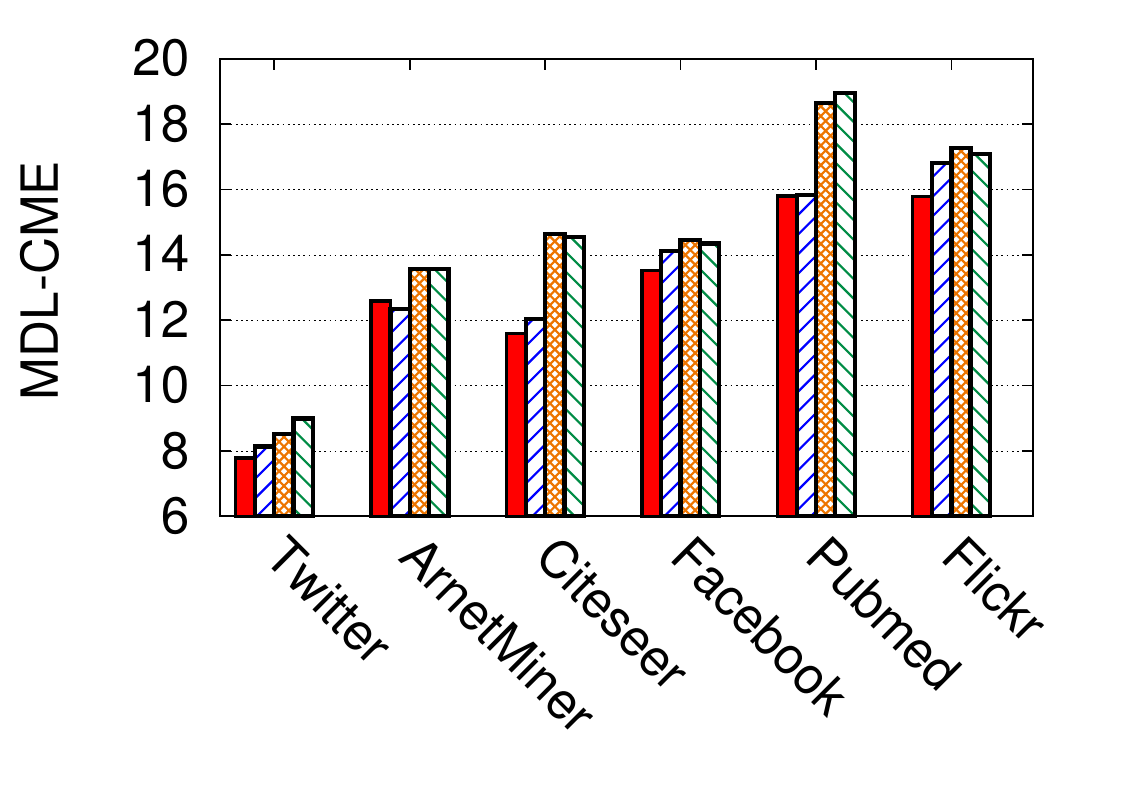}}
\caption{Comparison of the impact of different initializations on (a)  F measure and (b) Minimum description length or the number of bits required by the Content Map Equation to describe the network partition.}\label{fig:InitPerformance}
\end{figure}

Having established that the top-down method gives a good trade-off between partition quality and runtime, we now investigate the effect of different optimizations of the top-down algorithm. Specifically, we look at the effect of the initialization, i.e., the initial assignment of nodes to modules (see Section~\ref{subsec:topdown}). We investigate whether leveraging attributes or links helps identify better modules. The intuition is that once the nodes are assigned to modules based on their attributes, CME can use information in the links to find a locally better solution. We use a topic modeling technique, e.g., LDA~\cite{BleiNJ03}, to make the initial assignment. LDA requires the number of topics to be specified; hence, LDA20 means that the number of topics was set to 20, and LDA means that the number of topics was set to the true number of classes in the respective dataset. Alternatively, we can initialize the partition based on links alone, e.g., using the Map Equation, and then use attributes information to find a locally better solution with CME. We compare the partition quality resulting from random initialization to that resulting from LDA or ME initializations.

Figure~\ref{fig:InitFmeasure} reports the F-measure of the partition identified by the top-down method using different initializations (purity and accuracy results are similar). Surprisingly, the results demonstrate that neither LDA nor ME initializations help much in terms of partition quality improvement. Random initializations achieve better F-measure scores than LDA in 5 of 6 datasets, and better than ME initialization in 3 of 6 datasets. Since the Content Map Equation already incorporates content information equally with link information, the LDA/ME initializations only reweigh (or increase) the contribution of content/link information, which deteriorates performance.

Finally, we look at the effectiveness of different initialization methods to compress a random walk on a content-rich network. The results, shown in Figure~\ref{fig:Mlength}, suggest that  both LDA and ME initializations generally do not lead to better compression. Since both ME and LDA initialization are very time-consuming (see Figure~\ref{fig:time}), it is better to use random initialization in the top-down search method.

\section{Conclusion}
We have proposed and evaluated an information theoretic meth\-od for finding the modular structure of networks with node attributes.
% We accomplished this by adding a term to the Map Equation of Rosvall \& Bergstrom~\cite{Rosvall08} that summarized
{Building on the Map Equation of Rosvall \&
Bergstrom~\cite{Rosvall08}, we incorporate a new term that summarizes}
the contribution of the attributes to the description length of a random walk.  By minimizing the resulting Content Map Equation, we are able to identify modules with a larger information flow among the nodes, where the nodes also have similar attributes.

Accounting for node attributes changes the discovered modules. Our empirical evaluation of several large real-world networks demonstrates that the Content Map Equation results in a partition that is closer to the ground truth division then using links alone, or using alternative methods that take attributes into consideration. Moreover, in contrast to other methods, our framework does not require ad-hoc parameters that control the contribution of links and attributes to structure. One drawback of the approach is that it does not capture the dependencies among attributes in module dictionaries.  Because partitioning results are insensitive to, e.g., duplication of attributes in a representation, any additional information supplied by highly correlated attributes is essentially ignored.
%For example, synonyms are treated as distinct features, even though they will be highly correlated in their occurrence.
It would be an interesting challenge to extend the information theoretic framework to take these dependencies into account.

%We have also been able to apply this concept to networks of epidemic processes.  In particular, we determined the description length of a random walk reweighted adjacency matrix, where the edge weights are taken to be the product of the incident nodes' eigenvector centrality.  This methodology produced results similar to that of spectral clustering on the reweighted graph, giving preference to cliques and clique-like structures.

\subsection*{Acknowledgments}
%This paper is based on work funded by the Air Force Office of Scientific Research under contracts FA9550-10-1-0569  and FA9550-10-1-0102, the National Science Foundation under grant 0915678, and the Department of Energy Office of Science ASCR Program in Applied Mathematics.
\small{
\bibliographystyle{abbrv}
\bibliography{references,lerman}

\begin{thebibliography}{10}

\bibitem{SDM12}
L.~Akoglu, H.~Tong, B.~Meeder, and C.~Faloutsos.
\newblock Pics: Parameter-free identification of cohesive subgroups in large
  attributed graphs.
\newblock In {\em SDM}, pages 439--450. SIAM, 2012.

\bibitem{BleiNJ03}
D.~M. Blei, A.~Y. Ng, and M.~I. Jordan.
\newblock Latent dirichlet allocation.
\newblock {\em Journal of Machine Learning Research}, 3:993--1022, 2003.

\bibitem{PageRank}
S.~Brin and L.~Page.
\newblock The anatomy of a large-scale hypertextual web search engine.
\newblock {\em Computer Networks and ISDN Systems}, 30:107--117, 1998.

\bibitem{nus-wide-civr09}
T.-S. Chua, J.~Tang, R.~Hong, H.~Li, Z.~Luo, and Y.-T. Zheng.
\newblock Nus-wide: A real-world web image database from national university of
  singapore.
\newblock In {\em CIVR}, 2009.

\bibitem{Chung:Spectral:97}
F.~R.~K. Chung.
\newblock {\em Spectral Graph Theory}, volume~92 of {\em CBMS Regional
  Conference Series in Mathematics}.
\newblock American Mathematical Society, Feb. 1996.

\bibitem{Entropypartition}
J.~D. Cruz, C.~Bothorel, and F.~Poulet.
\newblock Entropy based community detection in augmented social networks.
\newblock In {\em CASoN}, pages 163--168. IEEE, 2011.

\bibitem{pascal-voc-2011}
M.~Everingham, L.~Van~Gool, C.~K.~I. Williams, J.~Winn, and A.~Zisserman.
\newblock The {PASCAL} {V}isual {O}bject {C}lasses {C}hallenge 2011 {(VOC2011)}
  {R}esults.

\bibitem{Fortunato10}
S.~Fortunato.
\newblock Community detection in graphs.
\newblock {\em Physics Reports}, 486:75--174, Jan. 2010.

\bibitem{FuICML09}
W.~Fu, L.~Song, and E.~P. Xing.
\newblock Dynamic mixed membership blockmodel for evolving networks.
\newblock In {\em Proceedings of the 26th Annual International Conference on
  Machine Learning}, pages 329--336. ACM, 2009.

\bibitem{HendersonEPF10}
K.~Henderson, T.~Eliassi-Rad, S.~Papadimitriou, and C.~Faloutsos.
\newblock {HCDF}: A hybrid community discovery framework.
\newblock In {\em SDM}, pages 754--7--65, 2010.

\bibitem{Imageclef10}
T.~D. Henning~Müller, Paul~Clough and B.~C. (Eds.).
\newblock {\em Experimental Evaluation in Visual Information Retrieval}.
\newblock The Information Retrieval Series, Vol. 32, Springer, 2010.

\bibitem{Huffman}
D.~Huffman.
\newblock A method for the construction of minimum-redundancy codes.
\newblock {\em Proc. Inst. Radio Eng.}, 40(9):1098--1101, September 1952.

\bibitem{huiskes08}
M.~J. Huiskes and M.~S. Lew.
\newblock The mir flickr retrieval evaluation.
\newblock In {\em MIR}, 2008.

\bibitem{Kang12homophily}
J.-H. Kang and K.~Lerman.
\newblock Using lists to measure homophily on twitter.
\newblock In {\em AAAI workshop on Intelligent Techniques for Web
  Personalization and Recommendation}, July 2012.

\bibitem{LongICML2006}
B.~Long, Z.~M. Zhang, X.~W\'{u}, and P.~S. Yu.
\newblock Spectral clustering for multi-type relational data.
\newblock In {\em Proceedings of the 23rd International Conference on Machine
  Learning}, pages 585--592. ACM, 2006.

\bibitem{Facebook}
J.~McAuley and J.~Leskovec.
\newblock Learning to discover social circles in ego networks.
\newblock {\em NIPS}, 2012.

\bibitem{conf/eccv/McAuleyL12}
J.~J. McAuley and J.~Leskovec.
\newblock In {\em ECCV (4)}, Lecture Notes in Computer Science, pages 828--841.
  Springer.

\bibitem{homophily}
M.~Mcpherson, L.~Smith-Lovin, and J.~M. Cook.
\newblock Birds of a feather: Homophily in social networks.
\newblock {\em Annual Review of Sociology}, 27:415--444, 2001.

\bibitem{Newman2006}
M.~E.~J. Newman.
\newblock Finding community structer in networks using the eigenvectors of
  matrices.
\newblock {\em Physical Review E}, 74(3), 2006.

\bibitem{icde/QiAH12}
G.-J. Qi, C.~C. Aggarwal, and T.~S. Huang.
\newblock Community detection with edge content in social media networks.
\newblock In {\em ICDE}, pages 534--545, 2012.

\bibitem{Ravasz2002Hierarchical}
E.~Ravasz, A.~L. Somera, D.~A. Mongru, Z.~N. Oltvai, and A.~L. Barab\'{a}si.
\newblock {Hierarchical Organization of Modularity in Metabolic Networks}.
\newblock {\em Science}, 297(5586):1551--1555, 2002.

\bibitem{Rives03}
A.~W. Rives and T.~Galitski.
\newblock Modular organization of cellular networks.
\newblock {\em Proc Natl Acad Sci U S A}, 100(3):1128--1133, 2003.

\bibitem{Rosvall08}
M.~Rosvall and C.~T. Bergstrom.
\newblock Maps of random walks on complex networks reveal community structure.
\newblock {\em Proceedings of the National Academy of Sciences},
  105(4):1118--1123, Jan. 2008.

\bibitem{wwwRuanFP13}
Y.~Ruan, D.~Fuhry, and S.~Parthasarathy.
\newblock Efficient community detection in large networks using content and
  links.
\newblock In {\em WWW}, pages 1089--1098, 2013.

\bibitem{sen:aimag08}
P.~Sen, G.~M. Namata, M.~Bilgic, L.~Getoor, B.~Gallagher, and T.~Eliassi-Rad.
\newblock Collective classification in network data.
\newblock {\em AI Magazine}, 29(3):93--106, 2008.

\bibitem{Shannon}
C.~Shannon.
\newblock A mathematical theory of communication.
\newblock {\em The Bell System Technical Journal}, 27:379--423, 1948.

\bibitem{Smith13socialcom}
L.~M. Smith, L.~Zhu, K.~Lerman, and Z.~Kozareva.
\newblock The role of social media in the discussion of controversial topics.
\newblock In {\em ASE/IEEE International Conference on Social Computing}, 2013.

\bibitem{Spielman07}
D.~A. Spielman and S.-H. Teng.
\newblock Spectral partitioning works: Planar graphs and finite element meshes.
\newblock {\em Linear Algebra and its Applications}, 421(2-3):284--305, Mar.
  2007.

\bibitem{ArnetMiner}
J.~Tang, J.~Sun, C.~Wang, and Z.~Yang.
\newblock Social influence analysis in large-scale networks.
\newblock {\em In Proceedings of the Fifteenth ACM SIGKDD International
  Conference on Knowledge Discovery and Data Mining}, 2009.

\bibitem{spectral-tutorial}
U.~von Luxburg.
\newblock A tutorial on spectral clustering.
\newblock {\em Statistics and Computing}, 17(4):395--416, Dec. 2007.

\bibitem{snobecluster}
C.~S. Wallace and D.~M. Boulton.
\newblock An information measure for classification.
\newblock {\em Computer Journal}, 11(2), 1968.

\bibitem{BACG12}
Z.~Xu, Y.~Ke, Y.~Wang, H.~Cheng, and J.~Cheng.
\newblock A model-based approach to attributed graph clustering.
\newblock In {\em SIGMOD Conference}, pages 505--516, 2012.

\bibitem{CESNAICDM13}
J.~Yang, J.~McAuley, and J.~Leskovec.
\newblock Community detection in networks with node attributes.
\newblock In {\em International Conference On Data Mining (ICDM)}. IEEE, 2013.

\bibitem{YangKDD09}
T.~Yang, R.~Jin, Y.~Chi, and S.~Zhu.
\newblock Combining link and content for community detection: a discriminative
  approach.
\newblock In {\em KDD}, pages 927--936, New York, NY, USA, 2009. ACM.

\bibitem{zhouCY09}
Y.~Zhou, H.~Cheng, and J.~X. Yu.
\newblock Graph clustering based on structural/attribute similarities.
\newblock {\em PVLDB}, 2(1):718--729, 2009.

\bibitem{ZhuNC11}
L.~Zhu, W.~K. Ng, and J.~Cheng.
\newblock Structure and attribute index for approximate graph matching in large
  graphs.
\newblock {\em Inf. Syst.}, 36(6):958--972, 2011.

\bibitem{getoor13}
Y.~Zhu, X.~Yan, L.~Getoor, and C.~Moore.
\newblock Scalable text and link analysis with {Mixed-Topic} link models.
\newblock In {\em Proc. of KDD}, 2013.

\end{thebibliography}

} 
\balancecolumns
% That's all folks!
\end{document}